\documentclass[12pt]{article}
\usepackage{graphicx}

\def\b{\beta}
\def\c{\chi}
\def\d{\delta}
\def\e{\epsilon}

\def\k{\kappa}
\def\l{\lambda}
\def\m{\mu}

\def\p{\pi}

\def\r{\rho}
\def\s{\sigma}

\def\x{\xi}

\def\S{\Sigma}

\def\vf{\varphi}

\def\beq{\begin{equation}}
\def\eeq{\end{equation}}
\def\bea{\begin{eqnarray}}
\def\eea{\end{eqnarray}}

\def\pl#1#2#3{Phys.~Lett.~{\bf B {#1}} ({#2}) #3}
\def\np#1#2#3{Nucl.~Phys.~{\bf B {#1}} ({#2}) #3}
\def\prl#1#2#3{Phys.~Rev.~Lett.~{\bf #1} ({#2}) #3}
\def\pr#1#2#3{Phys.~Rev.~{\bf D {#1}} ({#2}) #3}

\newcommand{\mpl}{M_{\rm P}}
\newcommand{\VEV}[1]{\langle #1 \rangle}

\newcommand\gev{\,\mbox{GeV}}

\newcommand\ha{\,\mbox{H}^{(1)}_{3/2}}
\newcommand\ai{\,\mbox{Ai}}
\newcommand\bi{\,\mbox{Bi}}

\renewcommand{\(}{\left(}
\renewcommand{\)}{\right)}
\renewcommand{\[}{\left[}
\renewcommand{\]}{\right]}

\begin{document}
\date{\mbox{ }}
\title{{\normalsize DESY 01-019\hfill\mbox{}\\
April 2001\hfill\mbox{}}\\
\vspace{2cm} \textbf{False Vacuum Decay after Inflation}\\
[8mm]}
\author{T.~Asaka, W.~Buchm\"uller, L.~Covi\\
\textit{Deutsches Elektronen-Synchrotron DESY, Hamburg, Germany}}
\maketitle

\thispagestyle{empty}

\begin{abstract}
\noindent
Inflation is terminated by a non-equilibrium process which finally leads to a 
thermal state. We study the onset of this transition in a class of hybrid 
inflation models. The exponential growth of tachyonic modes leads to 
decoherence and spinodal decomposition. We compute the decoherence time, 
the spinodal time, the size of the formed domains and the homogeneous 
classical fields within a single domain.
\end{abstract}

\newpage

An inflationary phase in the early history of the universe can explain its 
present flatness and homogeneity as well as the anisotropy of the cosmic 
microwave background \cite{lin90}. Inflation has to terminate
in some non-equilibrium process which eventually leads to a thermal state. 
This is elegantly realized in hybrid models of inflation where the time 
evolution of the inflaton field $\vf$ controls the transition from a false 
vacuum state to the true vacuum \cite{lin91}. 
Hybrid inflation arises naturally in supersymmetric
theories with spontaneous symmetry breaking \cite{clx94}. The dynamics
of the inflaton field during the inflationary phase is then determined
either by supergravity corrections \cite{clx94},
by quantum corrections \cite{dss94} or by the effects of the
supersymmetry breaking sector of the theory \cite{bcd00}.

The phase transition at the end of inflation is of crucial importance
since it determines the initial conditions for the further evolution towards
the thermal state. Important quantities, which have to be obtained, are
the reheating temperature and the initial
abundances of weakly coupled particles, such as gravitinos, which are not
in thermal equilibrium. In hybrid inflation the decay of the 
false vacuum takes place via spinodal decomposition \cite{gss83}, 
similar to a second-order phase transition. Such a decay has already been 
discussed in the literature in the context of new inflation \cite{gp85,bvx98} 
and also for scalar $\phi^4$-theory in Minkowsky space \cite{fgx00}. 

In the following we shall study the onset of this transition in a
class of hybrid inflation models \cite{bcd00}, where the
evolution is controlled by the slow-roll motion of the inflaton field. 
This will allow us to follow in detail the breakup of the homogeneous
inflationary phase into domains and
to compute the decoherence time, the spinodal time, the average size of the 
formed domains and the homogeneous classical fields inside a single domain. \\

\noindent\textbf{False vacuum inflation}\\

Hybrid inflation can be based on the superpotential \cite{bcd00}
\bea
W = W_G + W_S 
= \lambda T \(M_G^2-\Sigma^2\)+ M_S^2 \(\beta + S\)\;.
\label{wgs}
\eea
Here $W_G$ describes the supersymmetric breaking of a global symmetry, whereas
the Polonyi superpotential $W_S$ accounts for supersymmetry breaking 
\cite{pol77}. The discrete $Z_2$ symmetry of the model leads to the well known
domain wall problem \cite{zko75}. We consider this case only for simplicity;
the extension to a continuous $U(1)$ symmetry, where no such problems
exist, is straightforward.
The mass $M_S$ is the scale of supersymmetry breaking, associated with the
gravitino mass $m_{3/2} = M_S^2/\mpl \exp (2-\sqrt{3})$, and the constant
$\beta \simeq (2-\sqrt{3}) \mpl$, where 
$\mpl=(8\p G_N)^{-1/2}=2.4\times 10^{18}\gev$ is the Planck mass. 
Successful inflation requires $M_S^2 \ll \l M_G^2$. A rotation
of the fields $T$ and $S$ by the small angle $\x \simeq M_S^2/(\l M_G^2)$
changes the superpotential (\ref{wgs}) to the standard form of an
O'Raifeartaigh model \cite{ora75}.

In the case of global supersymmetry one obtains from the superpotential
(\ref{wgs}) the scalar potential 
\bea
V_0 = \l^2 |M_G^2 - \S^2|^2 + 4\l^2 |T|^2 |\S|^2 + M_S^4\;,
\eea
with the ground state 
\bea
\VEV{T} = 0\;, \quad |\VEV{\S}| = M_G\;.
\eea
For $|T|^2 > M_G/2$, $V_0$ has a local minimum at $\S = 0$ where it is flat 
in $S$ and $T$.

Supergravity corrections are included in the potential
\beq\label{vsugra}
V = e^{K/\mpl^2} 
\[\left|{\partial W \over \partial z_i} + {z_i^* W\over \mpl^2} 
\right|^2 - 3 
{|W|^2 \over \mpl^2} \]\;,
\eeq
where $z_i = T, \S, S$, $i=1\ldots 3$. We choose  $K$  to be the 
canonical K\"ahler potential, $K = \sum_i |z_i|^2$, which is singled out
in $N=2$ supersymmetry \cite{wy01}. Up to corrections
${\cal O}(1/\mpl^n)$, the expectation values of the three fields $T$,
$\S$ and $S$ in the true vacuum are 
\beq
\VEV{T} = 0\;,\quad |\VEV{\Sigma}| = M_G\;, \quad \VEV{S}=(2-\sqrt{3})\mpl\;.
\eeq

The supergravity corrections distinguish 
between the real and the imaginary part of $T \simeq (\vf + i\c)/\sqrt{2}$.
For $\vf > \vf_c = M_G$, $V$ has again a local minimum at
$\S = (\s + i\r)/\sqrt{2} = 0$. The energy density is then 
$V \simeq \l^2 M_G^4$, which leads to an inflationary phase with Hubble
parameter $H = \l M_G^2/(\sqrt{3} \mpl)$. Here the potential is almost flat 
in $\vf$ which plays the role of the inflaton. The mass of $\r$ is large,
$m_\r \simeq 2\l \vf > 2\l M_G$. The minimum with respect to $S$ and $\c$
lies at $S=\c=0$, but the corresponding masses are small, 
$m_S = {\cal O}(H)$ and $m_\c = {\cal O}(H\vf/\mpl)$.

In this paper we are interested in the decay of the false vacuum driving
inflation, which takes place near the critical point $\vf_c = M_G$. This
phase transition is determined by the dynamics of the lightest scalar fields,
the inflaton $\vf$ and $\s$, which becomes massless at $\vf_c$. The scalar
potential for $\vf$ and $\s$ reads, up to corrections ${\cal O}(1/\mpl^2)$,
\bea
V(\vf,\s) = \l^2 \(M_G^2 - {\s^2\over 2}\)^2 + \l^2\vf^2\s^2 + \k\vf\;.
\eea
The linear term in the potential is ${\cal O}(1/\mpl)$, with
\bea
\k = 2^{3/2} (2-\sqrt{3})\,\l\, {M_S^2 M_G^2\over \mpl}\simeq 
0.6\, \l m_{3/2} M_G^2\;.
\eea

For a certain range of parameters the time evolution of the inflaton field
during inflation is determined by the linear term in the potential 
\cite{bcd00}. Representative values are $M_S \simeq 1.6\times 10^{10}\gev$,
which implies $m_{3/2} \simeq 130\gev$,  $\l\simeq 1.3 \times 10^{-5}$ and 
$M_G \simeq 10^{16}\gev$. This yields $H \simeq 3.1 \times 10^8\gev$ and
$\k \simeq (4.6\times 10^9 \gev)^3$. From the full supergravity potential
(\ref{vsugra}) one obtains for the slow-roll conditions of the inflaton $\vf$,
as long as $\s^2$ is sufficiently small,
\bea
\e &=& {\mpl^2\over 2} \({V'\over V}\)^2 = {4 \xi^2 \b^2\over \mpl^2}
\(1- {\vf^3\over 2\sqrt{2} \xi\b \mpl^2} + \ldots\)\ \ll 1\;, \\
\eta &=& \mpl^2 {V''\over V} = 
{3\over 2} {\vf^2\over \mpl^2} + \ldots\ \ll 1\;.
\eea
These conditions are clearly satisfied for values of $\vf \ll \mpl$ and
in particular below the critical value $\vf_c$. The solution of the equation 
of motion for the homogeneous classical field reads,
\bea\label{slow}
\vf_{cl}(t) = M_G - {\k\over 3H} t\;,
\eea
where we have fixed the origin of time by $\vf(0) = \vf_c = M_G$. The
transition from the inflationary phase to the thermal phase starts at $t=0$.\\

\noindent\textbf{Fluctuations near the critical point}\\

At $\vf = \vf_c$ the mass term of the $\s$-field changes sign. This is
analogous to a second-order phase transition with $\vf(t)$ playing the
role of a time dependent temperature of the thermal bath. In flat 
de Sitter space with metric
\bea
ds^2 = dt^2 - e^{2Ht} d\vec{x}^2\;,
\eea
the quantum fields $\vf(t,\vec{x})$ and $\s(t,\vec{x})$ can be expanded 
into plane waves which are momentum eigenstates,
\bea
\vf(t,\vec{x}) &=& \vf_{cl}(t) +
e^{-{3\over 2}Ht} \int{d^3k\over (2\pi)^{3/2}}\(
a_{\vf}(\vec{k}) \vf_k(t) e^{i\vec{k}\vec{x}} + 
a^\dag_{\vf}(\vec{k}) \vf^*_k(t) e^{-i\vec{k}\vec{x}}\)\;,\\
\s(t,\vec{x}) &=& 
e^{-{3\over 2}Ht} \int{d^3k\over (2\pi)^{3/2}}\(
a_{\s}(\vec{k}) \s_k(t) e^{i\vec{k}\vec{x}} + 
a^\dag_{\s}(\vec{k}) \s^*_k(t) e^{-i\vec{k}\vec{x}}\)\;.
\eea 
Note the absence of $\s_{cl}(t)$, the homogeneous part of $\s(t,\vec{x})$.
This is due to the initial condition $\s_{cl}(0)=0$ at the end of inflation. 
The annihilation and creation operators satisfy canonical commutation
relations,
\bea
[a_{\vf}(\vec{k}),a^\dag_{\vf}(\vec{k'})]=
[a_{\s}(\vec{k}),a^\dag_{\s}(\vec{k'})]=\d^3(\vec{k}-\vec{k'})\;,
\eea
and the canonically conjugate operators to the fields $\vf$ and $\s$ are
\bea
\pi_\vf(t,\vec{x}) = e^{3Ht} \dot{\vf}(t,\vec{x})\;,\quad
\pi_\s(t,\vec{x}) = e^{3Ht} \dot{\s}(t,\vec{x})\;.
\eea
The assumption of a constant Hubble parameter is well
justified during inflation and the onset of the subsequent phase transition.

Near the critical point $\vf = \vf_c$ the potential for the quantum fields
$\vf$ and $\s$ takes the form,
\bea
V(\vf,\s) = - {1\over 2} D^3 t\ \s^2 + \ldots \;.
\eea
where $D^3 = 4\l^2\k M_G/(3H)$. For the parameters given above one has
$ D=3.4 H$. Note, that this equation also holds in other inflationary models
where the inflaton is slowly rolling near $\vf = \vf_c$, since $D^3 t$ is
the first term in an expansion of the $\s$-mass around $t=0$.

The mode functions $\vf_k(t)$ and $\s_k(t)$ satisfy the 
equations of motion,
\bea
\ddot{\vf_k} + \(k^2 e^{-2Ht} - {9\over 4} H^2\)\vf_k = 0\;, \label{des}\\
\ddot{\s_k} + \(k^2 e^{-2Ht} - {9\over 4} H^2\ - D^3 t\)\s_k = 0\;.\label{nos}
\eea
The solution of eq.~(\ref{des}) is well known from studies of density
fluctuations during inflation. It is given by a Hankel function
\bea\label{han}
\vf_k(t) = {1\over 2} \({\pi\over H}\)^{1/2}\ \ha(z)\;, 
\eea
where $z=k\exp{(-Ht)}/H$ and
\bea
\ha(z) = - \({2\over \pi z}\)^{1/2}\ e^{iz}\ {z+i\over z}\;.
\eea

The solution of eq.~(\ref{nos}) has to be determined numerically. However,
asymptotically, for $t\ll t_k$ and $t\gg t_k$, where
\bea\label{match}
k^2 e^{-2Ht_k}-{9\over 4}H^2 = |D^3 t_k|\;,
\eea
useful approximate analytical solutions can be found\footnote{\noindent
For momenta $3H/2 < k <k_{max}\simeq (D^3/(2He))^{1/2} \simeq 3 H$ 
eq.~(\ref{match}) has more than one solution for $t_k$. We choose the smallest
value for $t_k$ to construct an approximate solution. For a more detailed
discussion, see \cite{abc}.}. At large negative
times we again have the Hankel function analogous to (\ref{han}) as solution.
This is the de Sitter-space initial condition. 

For $t\gg t_k$ eq.~(\ref{nos}) reduces to
\bea 
\ddot{\s_k} - D^3 t \s_k = 0\;.
\eea
which describes a scalar field in Minkowski space with mass squared 
$\m^2(t) = - D^3 t$, decreasing with time. The solutions are given
by the Airy functions $\ai$ and $\bi$ \cite{as64}. An approximate analytic
solution of eq.~(\ref{nos}) can be obtained by matching Airy and Hankel
functions at $t=t_k$,
\bea\label{air}
\s_k(t) = C_A(k) \ai(z) + C_B(k) \bi(z)\;,
\eea
where $z=D t$. For soft momenta, $k < H$, one finds
\bea
C_A= i\sqrt{{\pi\over 2D}}\;, \quad 
C_B= \sqrt{{\pi\over 2D}}\;.
\eea
For $z=D t\gg 1$ the behaviour of $\vf_k(t)$ is dominated by the 
asymptotic growth of $\bi(z)$,
\bea
\bi(z) \simeq {1\over \sqrt{\pi}} z^{-1/4} \exp{\({2\over 3} z^{3/2}\)}\;.
\eea
The physical meaning of such tachyonic instabilities was studied in
detail by Guth and Pi for fixed negative mass squared, $\m^2(t) = -const$, 
in the context of new inflation \cite{gp85}. More recently, the dynamics
of spinodal decomposition has been investigated by several groups
\cite{bvx98,fgx00}. 

The situation considered here is similar to the previously discussed cases
with respect to the occurrence of spinodal decomposition. However, a crucial
difference is that the negative curvature of the potential is neither 
constant nor suddenly `turned on'. The change from positive to negative
curvature is, on the contrary, controlled by the slow-roll of the inflaton
field across the critical point $\vf=\vf_c$.
\begin{figure}
\centering 
\includegraphics[scale=0.6]{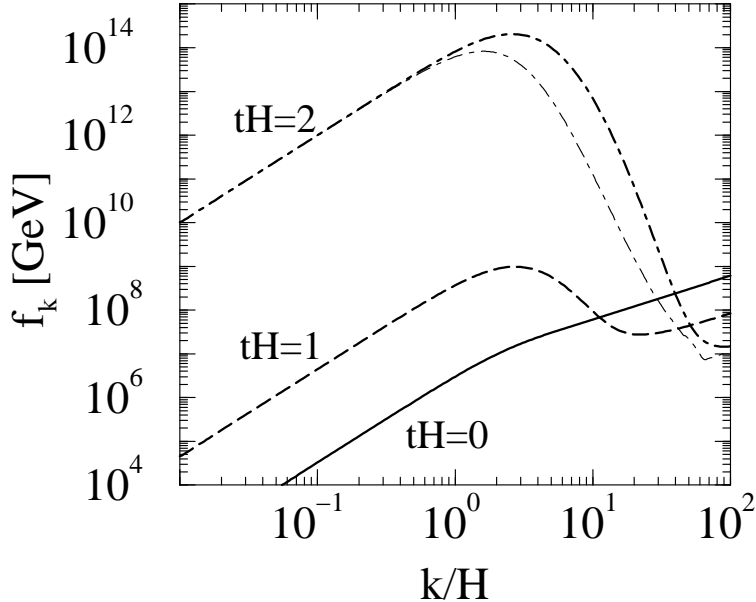}
\caption{{\it Spectral function $f_k(t)$ of the variance 
$\VEV{\s^2(t)}_r^{1/2}$. Thick lines correspond to numerical solutions; 
the thin line denotes an approximate analytical solution.}}
\end{figure}

An important quantity is the growth of the total fluctuation, the variance
\bea
\VEV{\s^2(t)}_r = \VEV{\s^2(t)} -  \VEV{\s^2(0)}\;,
\eea
where
\bea
\VEV{\s^2(t)} = \int {d^3k\over (2\pi)^3} e^{-3Ht} |\s_k(t)|^2
\equiv \int dk f_k(t)\;.
\eea
Results for the spectral function $f_k(t)$ are shown in Fig.~1. The expected
rapid growth with time is clearly visible. For small $k$ one has
$f_k \propto k^2$, since $|C_B(k)|\simeq const$. For larger momenta,
$k > k_* \sim 3 H$, also the time $t_k$ is large, which leads to a delayed 
growth of these modes and a rapid fall-off of $f_k(t)$ at large $k$.
For very large momenta, $k \gg k_*$, one sees the behaviour of a massless
field, i.e.  $f_k \propto k$, which is redshifted with increasing time 
due to the Hubble expansion.

The growth of the variance $\VEV{\s^2(t)}_r$ is shown 
in Fig.~2. It clearly exhibits the characteristic feature of our model, a
faster than exponential growth with time. Eventually, this growth is 
terminated by the back reaction of the produced fields. The spinodal time at 
which the formation of domains is completed can be estimated by comparing
the variance with the global minimum of the potential,
\bea
\VEV{\s^2(t_{sp})}_r \simeq 2 M_G^2\;. 
\eea
From Fig.~2 one reads off $t_{sp} \simeq 3/H$.
\begin{figure}
\centering 
\includegraphics[scale=0.6]{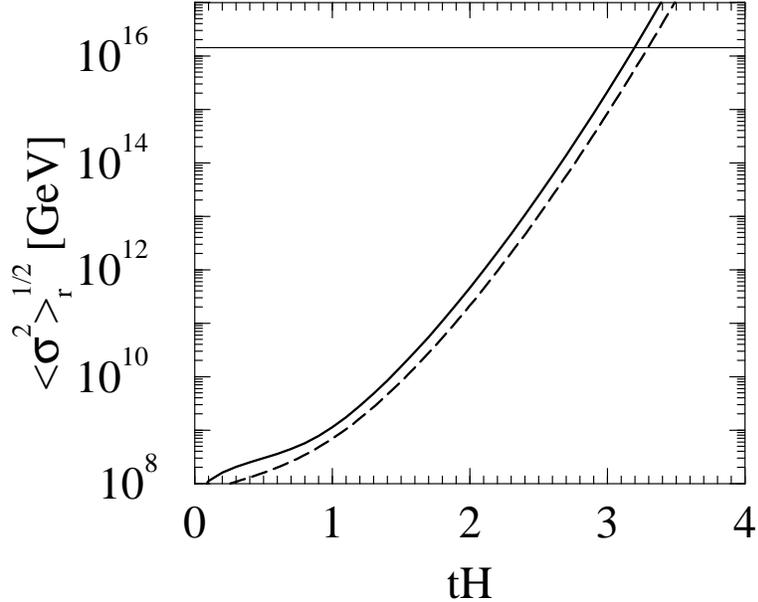}
\caption{{\it Growth of the variance $\VEV{\s^2(t)}_r$ with time. Full
and dashed lines correspond to numerical and analytical results, respectively.
The horizontal line denotes the true vacuum $\VEV{\s^2}_r =2 M_G^2$.}}
\end{figure}

Tachyonic modes $\s_k$ become classical at the
decoherence time $t_{dec}$. At later times the system is characterized by
a classical probability distribution $f(\s_k,t)$ for the amplitude $\s_k$. 
The homogeneous mode of the system $\s_{cl}(t) = \VEV{\s(t,\vec{x})}$ 
remains zero after spinodal decomposition \cite{ww87}. 

The decoherence time $t_{dec}$ is determined by the requirement that the
tachyonic modes become classical, i.e., the product $|\s_k(t)\pi_{\s k}(t)|$
has to be larger than $1/2$, the minimal value obtained for an oscillating
mode \cite{gp85}. For large times, $Dt\gg 1$, the product is determined by the 
exponentially growing Airy function $\bi(z)$,
\bea
|\s_k(t)\pi_{\s k}(t)| \simeq {1\over \pi} D |C_B(k)|^2
\exp{\({4\over 3}(Dt)^{3/2}\)}\;.
\eea
For soft modes $k < H$, the condition $|\s_k(t_{dec})\pi_{\s k}(t_{dec})|=
R_{dec} \gg 1$ yields the decoherence time 
\bea
t_{dec} \simeq {1\over D} \({3\over 4}\ln{(2R_{dec})}\)^{2/3}\;.
\eea
For $R_{dec}=100$ this implies $t_{dec}\sim 3/D \sim 1/H$. Results for the 
decoherence time obtained from the numerical integration are shown in Fig.~3.
\begin{figure}
\centering 
\includegraphics[scale=0.6]{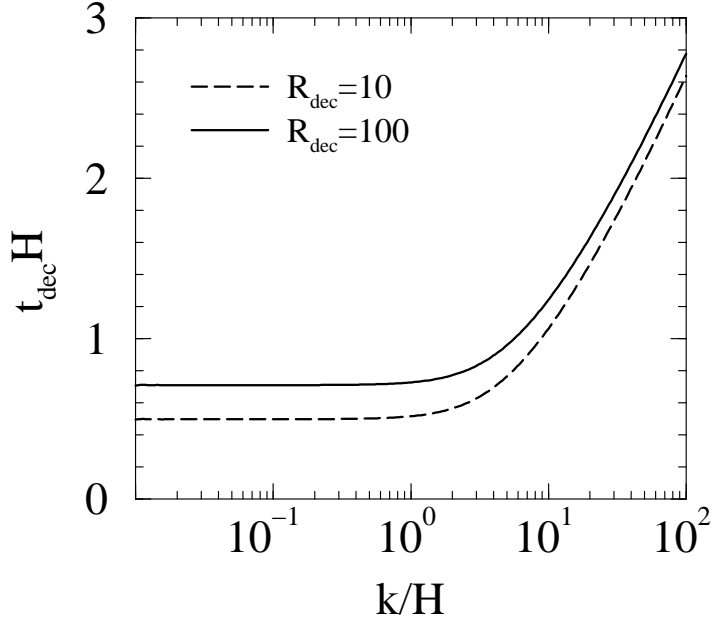}
\caption{{\it Decoherence time for tachyonic modes as function of momentum.}}
\end{figure}

From eq.~(\ref{slow}) and from  Fig.~3 one obtains the values of the
inflaton field $\vf$ and the variance $\VEV{\s^2}^{1/2}$ at the decoherence 
time $t_{dec} \sim 3/D$,
\bea
M_G - \vf_{cl}(t_{dec}) \sim 10^{12}\gev\;,\quad
\VEV{\s^2(t_{dec})}_r^{1/2} \sim 10^9 \gev\;.
\eea
These values can serve as initial conditions for the further time evolution
of the system \cite{kls97}. Note, that they are different for the two fields
and much larger than the frequently assumed initial condition $H/(2\pi)$.\\

\noindent\textbf{Formation of domains}\\

The rapid growth of tachyonic modes leads to a breakup of the homogeneous 
inflationary phase into domains. This spinodal decomposition is conveniently
described by means of fields averaged over a volume $V_l = l^3$ 
characterized by some smearing function \cite{gp85,ww87}. For simplicity
we shall use just a momentum cutoff ($k_l = 2\pi/l$),
\bea
\s_l(t,\vec{x}) = 
e^{-{3\over 2}Ht} \int_{k<k_l}{d^3k\over (2\pi)^{3/2}}\(
a_{\s}(\vec{k}) \s_k(t) e^{i\vec{k}\vec{x}} + 
a^\dag_{\s}(\vec{k}) \s^*_k(t) e^{-i\vec{k}\vec{x}}\)\;.
\eea  
One then obtains for the correlation function of the smeared operator 
($x=|\vec{x}|$),
\bea
\VEV{\s_l(t,\vec{x})\s_l(t,\vec{0})} =
{1\over 2\pi^2 x} e^{-3Ht} \int_{k<k_l} dk k |\s_k(t)|^2 \sin{kx}\;.
\eea
At time $t$ the physical distance and the physical momentum are given by
$y=x\exp{(Ht)}$ and $p=k\exp{(-Ht)}$, respectively. 
\begin{figure}
\centering 
\includegraphics[scale=0.6]{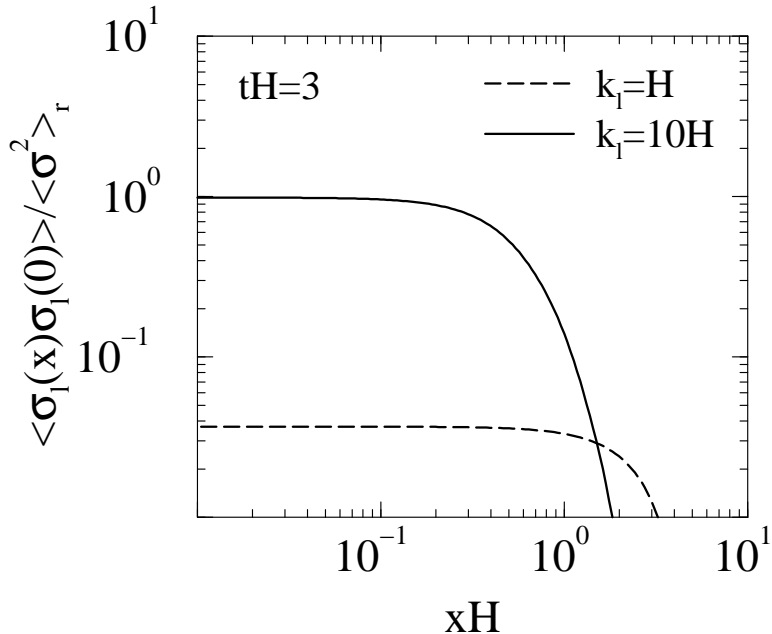}
\caption{{\it Correlation function of the smeared $\sigma$-field.}}
\end{figure}

An analytic estimate for the correlation function can be obtained 
based on the matching of Hankel and Airy functions
described above. Consider first the case of short times, $Dt<1$, and short 
distances, $1/k_l < x \ll 1/H$. Here one finds,
\bea
\VEV{\s_l(t,\vec{x})\s_l(t,\vec{0})} \simeq
- {1\over 4\pi^2 y^2} \(\cos{(k_l x)} - 1\)\;.
\eea
This is the result expected for a massless scalar field in Minkowsky space. 
The wavelength of the oscillation is given by the ultraviolet cutoff, 
$l = 2\pi/k_l$. 

For large times, $Dt \gg 1$, the correlation function is
dominated by the tachyonic modes with $k<k_*$. For distances $x < 1/k_*$
the deviation from homogeneity is small,
\bea
\VEV{\s_l(t,\vec{x})\s_l(t,\vec{0})} \simeq \VEV{\s^2(t)}_r
\( 1 + {\cal O}((k_* x)^2)\)\;.
\eea
For large distances, $x > 1/k_*$, the correlation function rapidly falls off.

Numerical results for the correlation function are shown in Fig.~4 for
two different values of the ultraviolet cut-off $k_l$. For a large cut-off,
$k_l = 10 H$, one reads off the correlation length 
$l_{corr} \simeq 1/(3 H) \simeq 1/k_*$, which corresponds to the average size 
of a domain. For cut-offs below $1/l_{corr}$, e.g. $k_l=H$, the expected 
decrease of the correlation function is observed. The energy density
stored in spatial variations of the $\s$-field is roughly given by
$\r_{grad} \sim M_G^2/l_{corr}^2$, which is much smaller than the total
energy density $V \simeq \l^2 M_G^4$. 

So far we have ignored all back reaction effects. They will terminate the
slow-roll motion of the inflaton field, damp the growth of the tachyonic
modes and modify the spinodal time $t_{sp}$. However, we do not expect
a change of our qualitative picture with small decoherence time, 
$t_{dec}<t_{sp}$, and asymmetric classical homogeneous fields $\vf$ and $\s$
larger than $H/(2\pi)$. The determination of the properties of the final 
thermal state, in particular the reheating temperature and the energy
fraction of topological defects, requires further work. 

Thanks to the slow-roll motion of the inflaton field $\vf$ across the critical
point $\vf_c$ we have been able to follow in detail the breakup of the
homogeneous inflationary phase into domains. The tachyonic modes of the
field $\s$, which become classical before the spinodal decomposition ends,
lead to an average domain size {\cal O}(1/H). Within a single domain they 
appear as a homogeneous classical field.

\end{document}